\documentclass[twocolumn,prl,aps,showpacs]{revtex4}

\usepackage[dvipdfm]{graphicx}

\begin{document}

\title{Nambu-Goldstone Mode in a Rotating Dilute Bose-Einstein Condensate}

\author{Masahito Ueda$^1$ and Tatsuya Nakajima$^2$}

\affiliation{
$^1$Department of Physics, Tokyo Institute of Technology, 
Meguro-ku, Tokyo 152-8551, Japan, \\ and
CREST, Japan Science and Technology Corporation (JST), Saitama 
332-0012, Japan \\
$^2$Physics Department, Graduate School of Science, 
Tohoku University, Sendai 980-8578, Japan}

\date{\today} 

\begin{abstract}
The Nambu-Goldstone mode, associated with vortex nucleation in a       
harmonically confined, two-dimensional dilute Bose-Einstein condensate,
is identified with the lowest-lying envelope of octupole-mode branches,
which are separated from each other by the admixture of quadrupolar    
excitations. As the vortex approaches the center of the condensate and 
the system's axisymmetry is restored, the Nambu-Goldstone mode becomes 
massive due to its coupling to higher rotational bands.
\end{abstract}


\maketitle 

One of the unique features of a gaseous Bose-Einstein condensate (BEC)  
is the fine tunability of interatomic interactions due to the Feshbach  
resonance~\cite{MIT,JILA}. 
This degree of freedom can be utilized to explore some unique features  
of a rotating BEC by making the strength of interaction close to zero;  
then as the angular momentum (AM), $L$, of the system increases, the    
low-lying states of the BEC in a harmonic potential become              
quasi-degenerate~\cite{WGS,Mottelson} and hence highly susceptible to   
symmetry-breaking perturbations. Such high susceptibility is considered 
to be the origin of vortex nucleation. Both experiments~\cite{Madison}  
and mean-field theories~\cite{Butts,Tsubota} have demonstrated that as  
$L$ increases, vortices enter the system from the outskirts of the BEC  
by spontaneously breaking the system's axisymmetry.
Associated with this symmetry breaking must be a Nambu-Goldstone mode   
(NGM) which, however, has been elusive.
This paper reveals the NGM using many-body theory and shows that this  
mode becomes massive as the vortex approaches the center of BEC at which
point the axisymmetry of the system is restored.

We consider a system of $N$ identical bosons, each with mass $M$, that  
undergo contact interactions and are confined in a two-dimensional      
harmonic potential with frequency $\omega$. 
Throughout this paper, we measure the length, energy, and AM in units  
of $(\hbar/M\omega)^{1/2}$, $\hbar\omega$, and $\hbar$, respectively.   
The single-particle and interaction Hamiltonians are then given by      
$H_0=\sum_{j=1}^N(-2\partial^2/\partial z_j\partial z_j^*+|z_j|^2/2)$   
and $V=2\pi g\sum_{j\neq k}\delta(z_j-z_k)$, where $z_j\equiv x_j+iy_j$ 
is the complex coordinate of the $j$-th particle and $g$ gives the      
ratio of the mean-field interaction energy per particle to $\hbar\omega$.
When $g\ll1$, it is legitimate to restrict the Hilbert space to that   
spanned by the basis functions 
$\phi_m(z)=(z^m/\sqrt{\pi m!}) \,e^{-|z|^2/2} \ (m=0,1,2,\cdots)$,      
where $m$ is the AM quantum number. In this ``lowest-Landau-level"      
approximation, the field operator is expanded as 
$\hat{\Psi}(z)=\sum_{m=0}^\infty \hat{b}_m \phi_m(z)$, 
where $\hat{b}_m$ is the annihilation operator of a boson with AM $m$.  
The second-quantized form of $V$ then reads~\cite{BP}
$\hat{V}=g\sum_{m_1,\cdots,m_4} V_{m_1,\cdots,m_4}\hat{b}_{m_1}^\dagger 
\hat{b}_{m_2}^\dagger \hat{b}_{m_3}\hat{b}_{m_4}$,
where 
$V_{m_1,\cdots,m_4}=\delta_{m_1+m_2,m_3+m_4}(m_1+m_2)!/
(2^{m_1+m_2}\sqrt{m_1!m_2!m_3!m_4!})$.
Since $\hat{H}_0=L+N$ is constant for given $L$ and $N$, 
the dynamics of the  
system is determined by $\hat{V}$ alone. 
The lowest-energy state of this system at a given $L$ is referred to as 
the yrast state, whose interaction energy is given for a repulsive case 
by $gN(N-1-L/2)$~\cite{BP,SW}. The trace of the yrast state viewed as a 
function of $L$ is called the yrast line.
In the following we will measure the energy of the system from that of  
the yrast state.

\begin{figure}[b]
\begin{center}
\includegraphics [width=.9\linewidth]{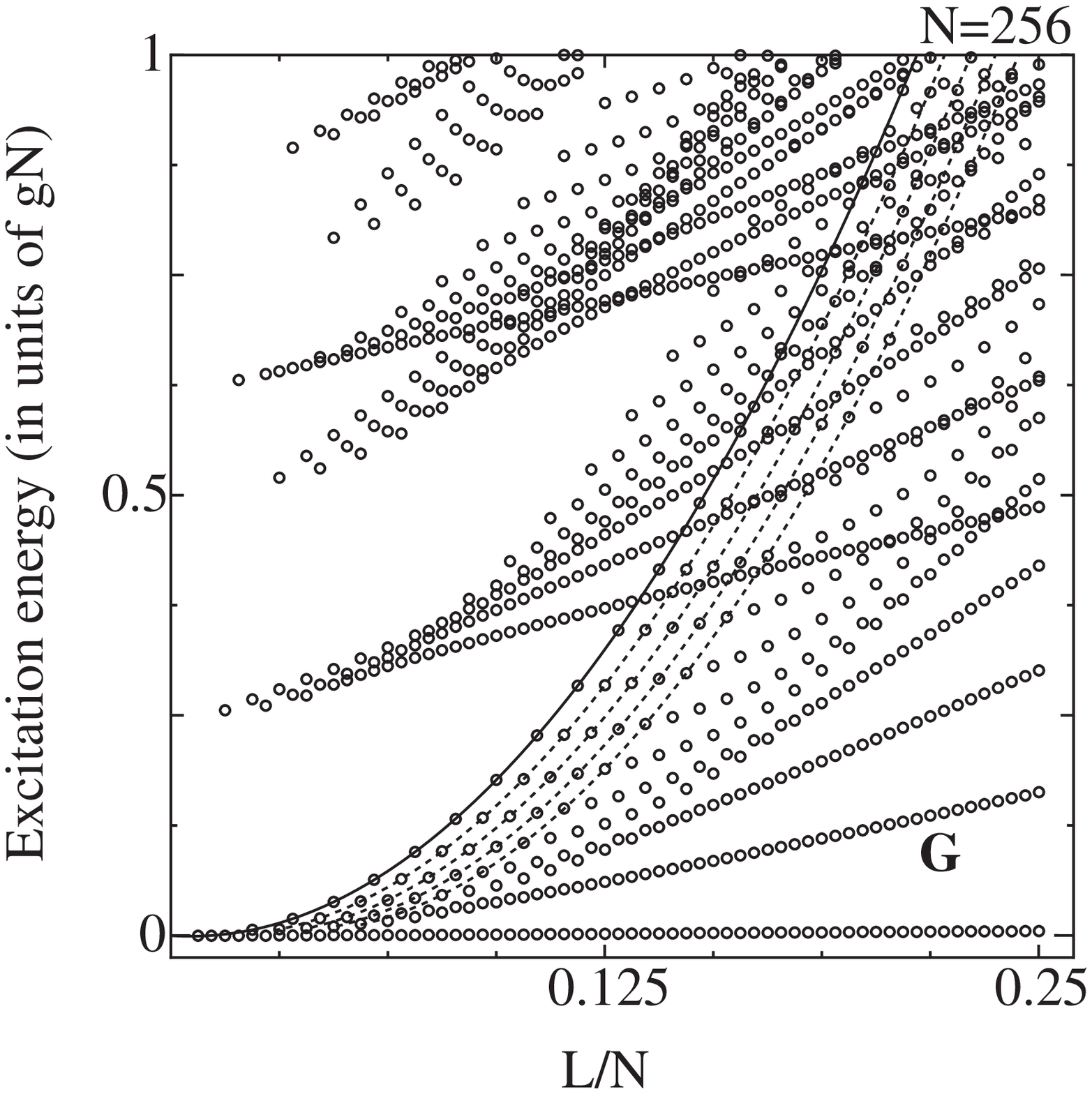}
\end{center}
\vspace*{-0.6cm}
\caption{
Nambu-Goldstone mode (labeled {\bf G}) and its mass acquisition in a 
rotating BEC.
The energy is measured from the yrast line (horizontal bottom line).
The solid curve shows the branch that arises from octupole-mode     
excitations alone and is described by $(27g/34)n_3(n_3-1)$, where   
$n_3$ denotes the number of octupole-mode excitations. The solid and
dotted curves are obtained from the least-squares fit of the data to
quadratic polynomials.
The dotted curves are displaced from the solid one by 2, 4, 6, 8    
units of angular momentum. These shifts are caused by the admixture 
of quadrupole-mode excitations. 
The lowest-energy envelope of these branches constitutes the        
Nambu-Goldstone mode which acquires mass as $L$ increases, as       
evident from the curve labeled {\bf G}. 
The excited states that involve the center-of-mass motion are not   
shown.
}
\label{f1}
\end{figure}

\noindent
{\it NGM in a rotating BEC}--.
The NGM should appear when the vortex is about to nucleate; that is,    
when the axisymmetry of the system is being broken.
In this regime ($L\ll N$), the excitation spectrum is divided into two  
groups having different characters~\cite{UN}.
One of them involves excitations whose energies are on the order of     
$gN$, and the other involves pairwise repulsive interactions between    
octupole modes with excitation energies given by
$\hat{O}=(27g/34)\hat{b}_3^\dagger\hat{b}_3(\hat{b}_3^\dagger\hat{b}_3-1)
$~\cite{UN}.
Because the energy scale ($\sim g$) of the latter group is smaller, by a
factor of $1/N$, than that of the former one and hence vanishes in the  
thermodynamic limit, one might suspect that $\hat{O}$ is the NGM.       
However, this is not the case.
We show that the NGM is the envelope of equally-spaced octupole-mode    
branches.
The envelope is labeled {\bf G} in Fig.~1, which is obtained by exact   
diagonalization of $\hat{V}$ for $N=256$.
This beautiful structure was not found previously because it emerges    
only for large $N$ and for relatively large values of $L/N$~\cite{Note}.
In Fig.~1, octupole-mode branches are indicated by solid and dotted     
curves, which are equally spaced with $\Delta L=2$. 
We have confirmed that this spacing is caused by the admixture of one   
quadrupole-mode excitation.

For a mode to qualify as a NGM, it must meet three conditions.
It must be massless, it must be associated with a broken symmetry,    
and it must play the role of restoring the broken symmetry.
Because the {\bf G} mode belongs to the second group 
discussed above,  
the excitation energy vanishes in the thermodynamic limit. Thus the first
condition is met.
Since the yrast states for $L \le N$ are energetically degenerate
under a rotation with angular 
velocity $\Omega = \frac {\partial E_{\rm tot}}{\partial L}
=\omega (1-\frac{gN}{2})$,
an axisymmetry breaking 
is expected to occur even for an infinitesimal, anisotropic
symmetry-breaking perturbation.
In fact, as the vortex enters the system, 
the axisymmetry is broken as seen in mean-field density profiles 
in Fig. 2~(a).
Thus the second condition is met.
The symmetry-restoring force is a correction to the mean-field result, so
that many-body theory must be invoked to find out whether the third 
condition is met.
We note that in exact diagonalization calculations the single-particle   
density-distribution function, 
$\rho({\bf r})=\langle\hat{\Psi}^{\dagger}({\bf r}) \hat{\Psi}({\bf r})  
\rangle$, is isotropic [i.e., $\rho({\bf r})=\rho(|{\bf r}|)$] and is not
suitable for studying the problem of axisymmetry breaking.
The symmetry breaking can be studied with the conditional distribution  
function (CDF)~\cite{cpf} defined as
$\rho({\bf r};{\bf r}_0) = \frac{1}{\rho ({\bf r}_0)} \,\langle         
\hat{\Psi}^{\dagger} ({\bf r}) \hat{\Psi}^{\dagger} ({\bf r}_0)         
\hat{\Psi} ({\bf r}_0) \hat{\Psi} ({\bf r}) \rangle $,
where ${\bf r}_0$ is the position of a test particle.
In order to ensure that the back action of placing the test particle on  
the system is negligible, we take $|{\bf r}_0|=3$ which is well outside 
the condensate. We set ${\bf r}_0=(r_0, 0)$ without loss of generality. 
Expanding the CDF in terms of $\cos(m\theta)$, we have
\begin{eqnarray}
{\tilde \rho}({\bf r}; {\bf r}_0)  &\equiv& \frac {N}{N-1}\, 
\rho({\bf r}; {\bf r}_0)-\rho (|{\bf r}|) \nonumber \\
&=& \sum_{m=0}^\infty C_m (r;|{\bf r}_0|) \, 
\frac{\cos(m \theta)}{\sqrt{\pi}} ,
\label{eq4}
\end{eqnarray}
where $r$ and $\theta$ are the polar coordinates of ${\bf r}$.

\begin{figure}[b]
\begin{center}
\vspace*{0cm}

\includegraphics[width=1.0\linewidth,bb=0 0 509 862]{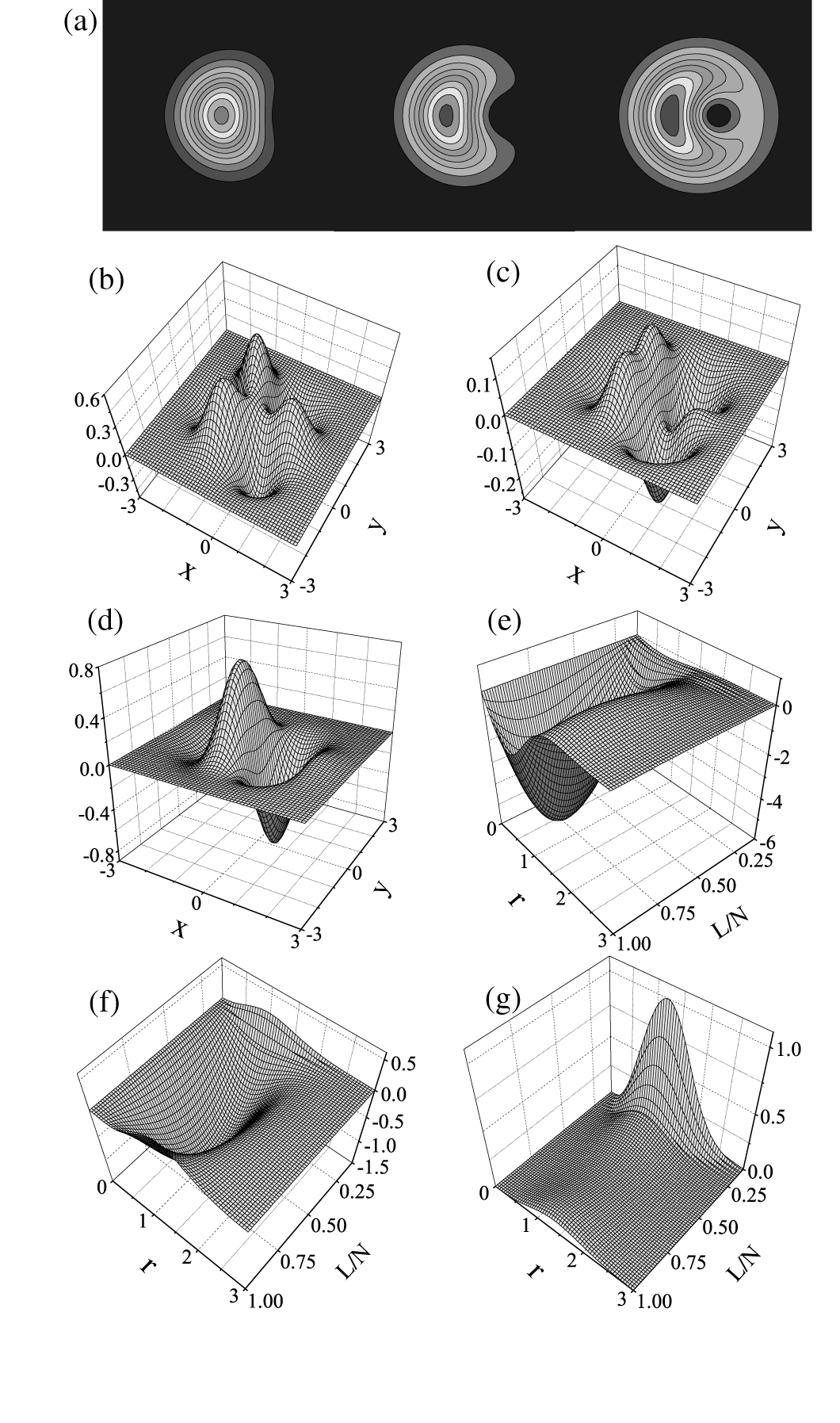}
\end{center}
\vspace*{-1.5cm}
\caption{
(a) Mean-field density profiles of BEC for $L/N=0.1, 0.4,$ and 0.8.
(b)-(c) Conditional distribution functions (CDFs) 
${\tilde \rho}({\bf r}; |{\bf r}_0|=3)$ defined in Eq.~(\ref{eq4}) 
for the quasi-degenerate lowest-lying excited state 
with $L=6$ (b) and $L=12$ (c). 
(d) CDF for the yrast state at $L=12$.
(e)-(g) show expansion coefficients $C_m (r;|{\bf r}_0|=3)$  
for $m=1,2$, and $3$, respectively, for the lowest-lying excited state 
for each $L$ satisfying $8 \le L \le N = 76$.
}
\label{f2}
\end{figure}

Figures 2(b) and (c) illustrate the CDF 
${\tilde \rho}({\bf r}; |{\bf r}_0|=3)$ 
for the lowest-lying excited states of 76 bosons with $L=6$ and $L=12$,  
respectively. 
The CDF in Fig.~2(b) features three-fold symmetry, reflecting the fact   
that the AM of this state is carried mostly by octupole-mode excitations.
We also note that one of the peaks is located near the boundary of the   
condensate where the vortex comes in. This implies that the octupole-mode
excitations compensate for the density depletion caused by the entrance  
of the vortex and thus plays the role of restoring the axisymmetry of the
system.
A further increase in $L$ is caused by quadrupole-mode excitations (see  
dotted curves in Fig. 1), so that the lowest-lying excited state         
gradually loses the character of three-fold symmetry, as illustrated in  
Fig.~2(c), and for larger $L$ the CDF eventually shows a dipole-like     
structure.
On the other hand, Fig.~2(d) shows the CDF of the yrast state at $L=12$, 
featuring a dipole-like structure.
The dipole-like distribution supports the entrance of the first vortex   
into the condensate by breaking the axisymmetry of the system.

Figures 2(e), (f), and (g) show the expansion coefficients               
$C_m (r;|{\bf r}_0|=3)$ in Eq.~(1) for $m=1$, 2, and 3, respectively, for
the lowest-lying excited states, which are constituting the {\bf G}      
branch in Fig.~1.
We note that $C_1 (r;|{\bf r}_0|=3)$ is negative with large magnitude    
around $r=1$ (i.e., around the periphery of the condensate), which       
suggests that this component is responsible for the invasion of the      
vortex.
On the other hand, the signs of $C_m (r;|{\bf r}_0|=3)$ for $m=2$ and    
$m=3$ are positive for $L \ll N$, and their magnitudes are maximal around
$r=1$.
This suggests that these two modes cooperate with each other to          
counteract the density depletion caused by the entrance of a vortex, in  
agreement with what is stated above.
We note that the amplitudes of $C_2$ and $C_3$ dwindle rapidly and that  
$C_2$ changes its sign when $L$ increases beyond the quasi-degenerate    
region. This, too, supports our claim that they constitute the NGM.      
We have confirmed that the signs of $C_4$ and $C_5$ are also positive,   
but that their magnitudes are negligible compared with that of $C_3$. 

\noindent
{\it Mass Acquisition of the NGM}--. 
Finally, we investigate the mechanism by which the NGM becomes massive as
$L$ increases. 
To this end, we separate the coordinates into center-of-mass (COM)       
$z_{\rm c}=(1/N)\sum_{j=1}^N z_j$ and relative ones                      
$u_j\equiv z_j-z_{\rm c}$, where $\sum _{i=1}^{N} u_i = 0$.
Accordingly, $H_0$ is decomposed into the COM part $H^{\rm com}=-(2/N)   
\partial^2/\partial z_{\rm c}\partial z_{\rm c}^* +N|z_{\rm c}|^2/2$     
and the relative part
$H^{\rm rel}= (2/N)\sum_{j,k=1}^{N-1}(1-N\delta_{jk})
\partial^2/\partial u_j\partial u_k^*+\sum_{j=1}^N|u_j|^2/2$.
The total Hamiltonian is thus decomposed into the COM part               
$H^{\rm com}$ and the relative part $H^{\rm rel}+V$, where               
$V=2\pi g\sum_{j\neq k}\delta(u_j-u_k)$. Therefore, a many-body wave     
function is separated into their counterparts:  
$\Psi(z_1,z_2,\cdots,z_N)=\Psi^{\rm com}(z_{\rm c})
\Psi^{\rm rel}(u_1,u_2,\cdots,u_{N-1})$.
In what follows, we will ignore the COM part because it is decoupled from
$V$.

We introduce a projection operator $e_i$, whose action is to eliminate  
those terms that contain the factor $u_i$ in the operand. For example,  
$e_iu_j=(1-\delta_{ij})u_j, \ e_iu_ju_k =(1-\delta_{ij})(1-\delta_{ik}) 
u_ju_k$,  etc. 
As shown below, many-body wave functions can be constructed in a        
systematic way by appropriate operation of $e_i$ on the yrast-state     
wave function, which is known for $0 \leq L \leq N$ to be               
$Y_L=\sum_{1\leq  i_1<\cdots<i_L\leq N}u_{i_1}\cdots u_{i_L}$~\cite{BP}.
The low-lying excitations of interacting Bose systems are collective in 
nature~\cite{Mottelson}; we claim that the main features (i.e., a large 
energy gap and almost linear rotational bands) of the excitation        
spectrum arise from a particular set of 
{\it coupled collective excitations} generated by operators 
$P_m=\sum_{i=1}^{N} u_i^me_i$ \ ($m=0,1,\cdots$).
These operators are analogous to Mottelson's multipolar operators       
$Q_m\propto \sum_iz_i^m$~\cite{Mottelson} but differ in that    
$z_i$ is replaced by $u_i$ and in that the projection operator $e_i$ is 
incorporated. Both of these modifications are essential for identifying 
an invariant subspace spanned by the yrast states and for constructing  
upon it the desired many-body excited states. 

We consider a linear superposition of the states 
$M_L^{(m)} \equiv P_mY_{L-m}$ \ ($m=3,4,\cdots,L$),
where $M_L^{(m)}$ describes an excitation in which one particle carries
AM $m$ and $L-m$ particles each carry a unit AM, for
\[
M_L^{(m)}=\sum_{i_1<\cdots<i_{L-m+1}}\!\!\!
\sum_{k=1}^{L-m+1}\!\!u_{i_k}^{m-1}\!\!
\prod_{l=1}^{L-m+1} \!\!\!u_{i_l} .
\]
The remaining $N-L+m-1$ particles carry no AM, and thus the condition  
$L\leq N+m-1$ must be met.

The transformation law of $M_L^{(m)}$ under the application of $V$ can 
be obtained as follows. 
The application of $2\pi\delta(u_i-u_j)$ on $M_L^{(m)}$ gives  
\begin{eqnarray}
& & \!\!\!\!\!\!\!\! 2\pi\delta(u_i-u_j)M_L^{(m)} 
= M_L^{(m)}+\frac{(u_i-u_j)^2}{4}e_ie_jM_{L-2}^{(m)}
\nonumber \\
& & \!\!\!\!\!\!\!\!
+ \left[ \frac{(u_i+u_j)^m}{2^{m-1}}-(u_i^m+u_j^m) \right] 
e_ie_jY_{L-m}
\nonumber \\
& & \!\!\!\!\!\!\!\!
+\left[\frac{(u_i+u_j)^{m+1}}{2^m}-(u_i^mu_j+u_iu_j^m)\right]
e_ie_jY_{L-m-1}.
\end{eqnarray}
Summing both sides of this equation over $i$ and $j(\neq i)$,
we obtain for $m\geq3$
\begin{eqnarray}
& & \!\!\!\!\tilde{V}M_L^{(m)}=
\left \{(m+2^{3-m}-4)N
+[2^{2-m}(m-1)+1]
\right.
\nonumber \\
& & \left.\!\!\!\!
\times (L-m)
+4(1-2^{1-m})\right \}M_L^{(m)}/2 -(1-2^{2-m})\nonumber \\
& &  \!\!\!\!\times(N-L+m)M_L^{(m+1)}/2+2^{2-m}m(L-m+1)
\nonumber \\
& & \!\!\!\!\!\times M_L^{(m-1)}
+R_{m+1}+\theta(m\geq4)R_m,
\label{GEN}
\end{eqnarray}
where $\tilde{V}\equiv V-N(N-1-L/2)$ and
\begin{eqnarray}
R_m=2^{1-m}\sum_{k=2}^{m-2} \frac{m!}{k!(m-k)!} P_{m-k}M_{L-m+k}^{(k)} . 
\label{R_m}
\end{eqnarray}
Equation~(\ref{GEN}), together with (\ref{R_m}), is the desired          
recursion relation. We see that each mode described by  $M^{(m)}_L$ is   
coupled with other modes $M^{(k)}_L$ with $k=3,4,\cdots,m-1$, and $k=m+1$.
Since the coefficient of $M^{(m+1)}_L$ in Eq.~(\ref{GEN}) includes a     
factor $N-L+m$ , the weight of this term decreases monotonically with    
increasing $L$. 
By introducing the truncation approximation discussed below and ignoring 
contributions from higher-order terms such as $M_l^{(l)}M_{L-l}^{(m)}$,  
we obtain a closed set of linear equations that can be solved easily even
for a large value of $m$. 
It turns out that to quantitatively reproduce the {\bf G} mode labeled in
Fig.~1, we must take into account $m$ up to a relatively large value,    
such as 7, because higher rotational bands couple strongly to 
the {\bf G} mode, as discussed below.
Writing Eq.~(\ref{GEN}) explicitly 
for $m=3, 4$, and $5$, 
we obtain
\begin{eqnarray}
\tilde{V}M_L^{(3)}&=&(L-\frac 94)M_L^{(3)}-\frac{N-L+3}{4}M_L^{(4)}
\nonumber \\
& & -\frac{3}{4} L(L-2) Y_L , \nonumber \\
\tilde{V}M_L^{(4)}&=&-\frac {L}{4}M_L^{(3)}+\frac{2N+7L-24}{8}M_L^{(4)} 
\nonumber \\ 
& &-\frac {3}{8} (N-L+4)M_L^{(5)}+\frac{3}{4} L(L-2) Y_L , \nonumber\\
\tilde{V}M_L^{(5)}&=&-\frac {5}{8}(L-3)M_L^{(3)}-\frac {5}{16}(L-2)M_L^{(4)}
\nonumber \\ 
& & 
+ \frac {10N+12L-55}{16} M_L^{(5)}
\nonumber \\
& & 
-\frac{7}{16} (N-L+5)  M_L^{(6)}, 
\label{spec}
\end{eqnarray}

\begin{figure}[t]
\begin{center}
\includegraphics[width=.9\linewidth]{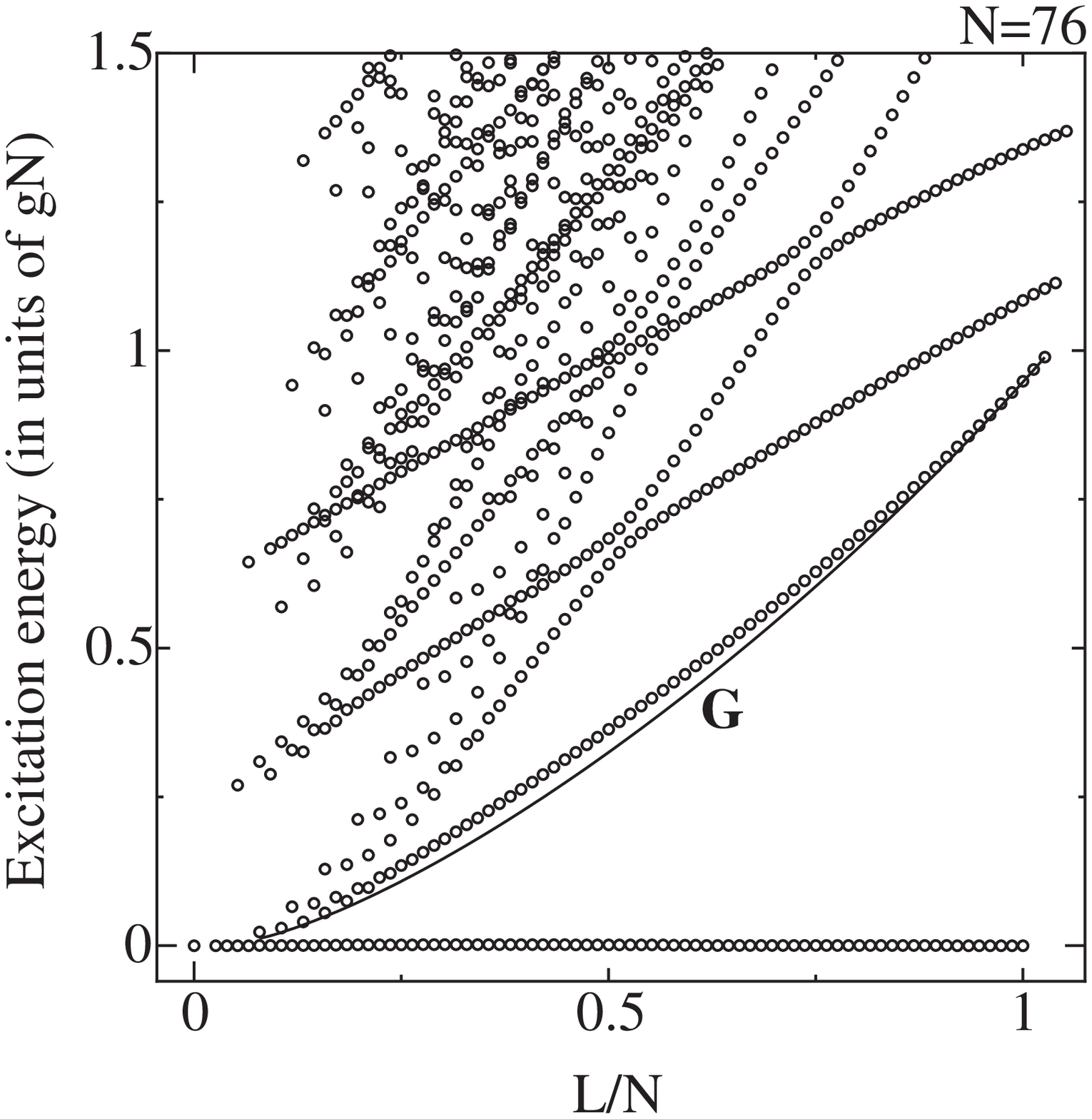}
\end{center}
\vspace*{-0.6cm}
\caption{
Comparison between the excitation energy of the {\bf G} mode 
obtained by solving a set of linearized equations 
for $m=3,4,\cdots,7$ (solid curve) and those    
obtained by the exact diagonalization of 
the Hamiltonian (open circles).
The energy is measured from the yrast state, which corresponds to   
the horizontal bottom line.
}
\label{f3}
\end{figure}

We note that the diagonal coefficients of $M_L^{(m)}$ in Eq.~(\ref{spec}),
are $(2N+7L)/8$ and $(5N+6L)/8$ for $m=4$ and $5$, respectively, and that 
they already reproduce rather well the almost linear spectra in           
Fig.~\ref{f3}. We refer to these as higher rotational bands.
We also note the coupling of $M_L^{(m)}$ ($m\ge4$) to $M_L^{(3)}$ becomes 
very strong, on the order of $O(N)$, near $L\sim N$.
To reproduce the lowest-lying branch consisting mainly of $M_L^{(3)}$, 
we have to solve the coupled equations for      
$m=3,4, \cdots$. 
The solid curve in Fig.~\ref{f3} shows the lowest excitation energy 
obtained by solving the above closed set of equations. 
It matches the {\bf G} mode rather well.

In conclusion, we investigated the Nambu-Goldstone mode (NGM) of a rotating
BEC associated with axisymmetry breaking due to vortex nucleation, and     
identified this mode with the lowest-lying envelope comprised of octupole  
branches that are equidistantly displaced by the admixture of quadrupole   
excitations.
We found that as the angular momentum of the system increases, the  
NGM acquires mass due to its strong coupling to higher rotational bands.

\begin{acknowledgments} 
 
M.U. and T.N. acknowledge support by Grants-in-Aid for Scientific 
Research (Grant Nos.~15340129 and 14740181) by the Ministry of Education, 
Culture, Sports, Science and Technology of Japan.

\end{acknowledgments}



\begin{thebibliography}{99}

\bibitem{MIT} 
S. Inouye, et al., Nature {\bf 392}, 151 (1998).

\bibitem{JILA} 
S.L. Cornish, et al., Phys. Rev. Lett. {\bf 85}, 1795 (2000).

\bibitem{WGS} N.K. Wilkin, et al.,
Phys. Rev. Lett. {\bf 80}, 2265 (1998).

\bibitem{Mottelson} B. Mottelson, 
Phys. Rev. Lett. {\bf 83}, 2695 (1999).

\bibitem{Madison}
K. W. Madison, et al.,
Phys. Rev. Lett. {\bf 84}, 806 (2000). 

\bibitem{Butts} D. A. Butts and D. S. Rokhsar, Nature {\bf 397}, 327 (1999).

\bibitem{Tsubota}
M. Tsubota, et al.,
Phys. Rev. A {\bf 65}, 023603 (2002).

\bibitem{BP} G.F. Bertsch and T. Papenbrock,
Phys. Rev. Lett. {\bf 83}, 5412 (1999).

\bibitem{SW}
R.A. Smith and N.K. Wilkin, 
Phys. Rev. A {\bf 62}, 61602 (2000);
T. Papenbrock and G.F. Bertsch, 
ibid. {\bf 63}, 023616 (2001).

\bibitem{UN}
T. Nakajima and M. Ueda, 
Phys. Rev. A {\bf 63}, 043610 (2001);
M. Ueda and T. Nakajima, 
ibid. {\bf 64}, 063609 (2001);
G.M. Kavoulakis, et al., 
ibid. {\bf 62}, 063605 (2000);
G.M. Kavoulakis, et al.,
ibid. {\bf 63}, 055602 (2001);
V. Bardek, et al., 
ibid. {\bf 64}, 015603 (2001); 
V. Bardek and S. Meljanac, 
ibid. {\bf 65}, 013602 (2002).

\bibitem{Note} In a previous 
paper [T. Nakajima and M. Ueda, Phys. Rev. Lett. {\bf 91}, 140401
(2003)], we reported the results of similar exact diagonalization
calculations for $N=25$. However, this size is too small 
to reveal a quasi-degenerate spectrum on the order of $g$ so that
we can distinguish the NGM from the yrast line. 
On the other hand, if we restrict $L$ as $L \ll N$, we can 
perform exact diagonalization calculations for a much larger
$N$, say, $N=10000$, but then we are not able to cover $L$ 
up to $L/N\sim 0.1$, which is necessary in order to investigate 
how the NGM becomes massive with increasing $L$.

\bibitem{cpf} 
X.J. Liu, et al., 
Phys. Rev. Lett. {\bf 87}, 030404 (2001).

\end{thebibliography}
\end{document}